\documentclass[aps,pra,twocolumn,superscriptaddress, nofootinbib]{revtex4-1}
\usepackage{graphicx}
\usepackage{amsmath}
\usepackage{subfigure}
\usepackage{braket}
\usepackage{color}

\begin{document}

\title{Selective reflection Casimir-Polder spectroscopy in vapor cells: the influence of 
the thermal velocity distribution}

\author{B. Dutta}
\affiliation{Laboratoire de Physique des Lasers, Universit{\'e} Sorbonne Paris Nord, 
F-93430, Villetaneuse, France}
\affiliation{CNRS, UMR 7538, LPL, 99 Avenue J.-B. Cl{\'e}ment, F-93430 Villetaneuse, 
France} 

\author{C. Boldt}
\affiliation{Institute of Physics, University of Rostock, Albert-Einstein-Stra{\ss}e 23-24, 
D-18059 Rostock, Germany}

\author{G. Garcia-Arellano}
\affiliation{Laboratoire de Physique des Lasers, Universit{\'e} Sorbonne Paris Nord, 
F-93430, Villetaneuse, France}
\affiliation{CNRS, UMR 7538, LPL, 99 Avenue J.-B. Cl{\'e}ment, F-93430 Villetaneuse, 
France}  

\author{M. Ducloy}
\affiliation{Laboratoire de Physique des Lasers, Universit{\'e} Sorbonne Paris Nord, 
F-93430, Villetaneuse, France}
\affiliation{CNRS, UMR 7538, LPL, 99 Avenue J.-B. Cl{\'e}ment, F-93430 Villetaneuse, 
France}

\author{P. Pedri}
\affiliation{Laboratoire de Physique des Lasers, Universit{\'e} Sorbonne Paris Nord, 
F-93430, Villetaneuse, France}
\affiliation{CNRS, UMR 7538, LPL, 99 Avenue J.-B. Cl{\'e}ment, F-93430 Villetaneuse, 
France} 

\author{S. Scheel}
\affiliation{Institute of Physics, University of Rostock, Albert-Einstein-Stra{\ss}e 23-24, 
D-18059 Rostock, Germany}

\author{A. Laliotis}
\email{laliotis@univ-paris13.fr}
\affiliation{Laboratoire de Physique des Lasers, Universit{\'e} Sorbonne Paris Nord, 
F-93430, Villetaneuse, France}
\affiliation{CNRS, UMR 7538, LPL, 99 Avenue J.-B. Cl{\'e}ment, F-93430 Villetaneuse, 
France} 

\date{July 8, 2025}

\begin{abstract}
 
Selective reflection is a high-resolution spectroscopic method that allows the probing of 
atomic and molecular gases in the near field of dielectric cell windows. 
It is a sensitive technique for measuring interactions between excited atoms and 
surfaces, thus providing a unique tool to probe resonant effects in Casimir-Polder physics. 
The theoretical modelling of selective reflection spectra has thus far employed
the infinite Doppler approximation which assumes that the Doppler shift dominates over 
the atom-surface interactions as well as the homogeneous transition linewidth. Here we 
extend the theoretical treatment of selective reflection spectra to include the effects of 
finite Doppler width. This allows us to analyse the regime of very large Casimir-Polder 
interactions, relevant for Rydberg atom spectroscopy, and revisit previous analyses of 
experiments with low-lying excited states. Further, this will aid in determining the 
signatures of quantum friction in vapor cell spectroscopy.  

\end{abstract}


\maketitle

\section{Introduction}
The Casimir-Polder (CP) interaction between atoms or molecules and surfaces is a paradigm 
of quantum electrodynamics with implications in fundamental precision measurements 
\cite{ballandPRL2024} and potential applications for probing electromagnetic properties of 
dielectrics \cite{carvalho_prl2023}. Casimir-Polder interactions have been probed with 
ground-state atoms in cold atom experiments 
\cite{cornell_PRA_2004, Bender_PRL_2010, cornell_PRL_2007} sensitive to the non-resonant 
part of the CP potential \cite{laliotisReview2022}. Resonant effects are accessible with 
excited atoms that can be probed via vapor cell spectroscopy. Resonant CP interactions 
allow one to probe the exotic regime of CP repulsion \cite{failacheprl1999}, thermal 
effects \cite{laliotisnatcommun2014}, and to tune CP interactions with meta-surfaces 
\cite{laliotis_scienceadv2018}. Rydberg-surface interactions are also ever-relevant in 
quantum physics \cite{SandoghdarPRL1992}, quantum technologies and hybrid systems 
\cite{kublernatphot2010}, and would allow fundamental CP tests of the extreme near field 
\cite{LaliotisPhysRevResearch2024}.   

Selective reflection (SR) and thin cell spectroscopy are important experimental methods 
that allow probing atomic 
\cite{Woerdman_1975,  akulshin_1982, laliotisReview2022, Meng_2016, Sargsyan:17} or 
molecular gases \cite{lukusa_prl_2021, arellano_probing_2024} in the nanometre or 
micrometre regime with sub-Doppler frequency resolution, in particular when a frequency 
modulation (FM) is applied \cite{akulshin_1982}. Frequency-modulated Selective Reflection 
(FMSR) has been extensively used for measuring Casimir-Polder interactions 
\cite{oriaepl1991, failacheprl1999, laliotisnatcommun2014, Carvalho_PRL_2023}. This is 
achieved by fitting the FMSR spectra with a theoretical model that accounts for transient 
effects that become important due to collisions of atoms with the hard surface of 
the cell window \cite{schuurmans_1976, DUCLOY1993336} and for near-field Casimir-Polder 
interactions that follow the inverse distance law, $-C_{3}/z^{3}$ 
\cite{ducloy_jphysique_1991}, where $C_3$ is the van der Waals coefficient and $z$ the 
atom-surface distance. The theoretical description of FMSR spectra (i.e the first 
derivative of SR spectra) \cite{ducloy_jphysique_1991} makes the explicit assumption of an 
infinite Doppler width (flat velocity distribution) that is justified as long as 
$ k v_p \gg \Gamma $, where $\Gamma$ is the homogeneous transition linewidth (including 
collisional broadening). It has been recognised, however, that the Doppler effect can 
influence the wings on sub-Doppler selective reflection spectra that could lead to 
systematic errors, especially for experiments on the D1 and D2 lines of alkali atoms. For 
this purpose, finite Doppler corrections are often included in the FMSR spectra 
\cite{Papag1994, laliotisAPB2008} but are calculated without accounting for Casimir-Polder 
interactions.

Although a full numerical calculation of SR spectra is challenging, it is important for 
Casimir-Polder experiments for the following reasons: (i) the 
interpretation of spectroscopic measurements with highly excited atoms (Rydberg atoms), 
where Casimir-Polder interactions are very large, the $C_3$ grows with the effective 
principal quantum number $n^{\star}$ as $(n^{\star})^4$; (ii) the evaluation of systematic 
effects that could have arisen during past measurements of the CP shift 
\cite{oriaepl1991, failacheprl1999, laliotisnatcommun2014} due to the infinite Doppler 
approximation; (iii) the evaluation of the 
spectroscopic signatures of quantum friction, a velocity-dependent atom-surface interaction 
that has remained experimentally elusive \cite{klattPRA2026, KlattPRA2022}. Atom-polariton 
coupling could boost the effects of quantum friction and lead to visible effects in 
selective reflection spectra \cite{klattPRA2026}, but a correct evaluation of these effects 
requires inclusion of the full velocity distribution of the atoms in the calculations.


Here we present a full numerical evaluation of the selective reflection spectra under the 
influence of Casimir-Polder interactions accounting for the Maxwell-Boltzmann (MB) distribution 
of the atomic velocities. We show that our approach is necessary for interpreting the large 
atom-surface interaction regime, by using the example of relatively small Rydberg states 
($n^{\star}\approx 15$) that are experimentally accessible in vapor cell experiments \cite{biplab_dutta_2023, Butery2025}. We also examine the systematic effects 
that may be introduced in Casimir-Polder measurements of low-lying states, such as the D1 
line of cesium \cite{laliotisAPB2008}. Finally we show that the infinite velocity 
distribution approximation is sufficient to account for the effects of a distance-dependent 
lifetime when atomic couplings are resonant with polariton resonances. 

\section{Selective Reflection spectroscopy}
For the calculation of selective reflection spectra, we assume a propagating plane wave 
inside a dielectric medium (the window of a vapor cell) at normal incidence on the 
window/vapor interface. The direction of propagation is $z$ and the frequency and 
wavevector component of the light are $\omega$ and $k$, respectively. We restrict our 
analysis to normal incidence as this realises linear sub-Doppler spectroscopy and 
corresponds to most experimentally relevant scenarios. In our calculations, we closely 
follow the approach of Ref.~\cite{ducloy_jphysique_1991}.

The reflectivity $r$ of the interface is defined by the ratio between reflected ($E_R$) 
and incident ($E_{in}$) electric fields as
\begin{equation}
\label{eq:reflectivity}
r = \frac{n-1}{n+1}-\frac{n}{(n+1)^2} \Bar{\chi} 
\end{equation}
where $\Bar{\chi}$ is the effective vapor susceptibility. It follows from 
Eq.~(\ref{eq:reflectivity}) that the reflection coefficient of the interface, $R=|r|^2$, is 
\begin{equation}
\label{eq:reflection}
R =  \left( \frac{n-1}{n+1} \right)^2-\frac{2n(n-1)}{(n+1)^3}\Re (\Bar{\chi}) \,. 
\end{equation}
The first term in Eq.~(\ref{eq:reflection}) is the non-resonant reflectivity in the absence 
of the vapor, which we denote by $R_o$, while the second term is the first-order correction 
due to the presence of the vapor, which is essentially due to the beating between the 
re-radiated atomic field and the non-resonant reflection off the window/vacuum interface. 
The second-order correction ($\propto\Bar{\chi}^2$) has been ignored. The effective 
susceptibility $\Bar{\chi}$ is given by 
\begin{equation}
    \Bar{\chi} = -\frac{2\mathrm{i}k}{\varepsilon_0 E_T} \int_0^\infty \mathrm{d}z\, p(z) \mathrm{e}^{2\mathrm{i}kz}
\end{equation}
where $\varepsilon_0$ is the vacuum susceptibility, $E_T$ the transmitted field and $p(z)$
is the macroscopic polarisation of the resonant (atomic or molecular) medium. The 
exponential $2kz$ denotes the extra phase of the reflected field re-radiated by atoms that 
are located at a position $z$ compared to the phase of the field re-radiated by atoms at 
$z=0$. Assuming that the vapor consists of two-level atoms ($\ket{g}$ and $\ket{e}$ for 
ground and excited states, respectively) and that the excitation field is weak, the 
polarisation of the vapor is given by
\begin{equation}
    p(z) = N\mu \int_{-\infty}^{\infty}\mathrm{d}^3\mathbf{v}\, W(\mathbf{v}) \sigma_{eg}(\mathbf{r},\mathbf{v})
\end{equation}
where $N$ is the atomic number density, $\mu$ the dipole moment matrix element of the 
atomic transition, $W(\mathbf{v})$ the MB velocity distribution of the atoms, 
and $\sigma_{eg}$ is the off-diagonal element of the density matrix
\begin{equation}
\label{eq:sigma_eg}
     \sigma_{eg} = \mathrm{i}\frac{\Omega}{2v_z} \int_{z_0}^{z} \mathrm{d}z'\, \mathrm{e}^{(\mathcal{L}(z')-\mathcal{L}(z))/v_z}.
\end{equation}
In Eq.~(\ref{eq:sigma_eg}), $z_o$ represents the distance after which the optical coherence 
vanishes, (i.e. $\sigma_{eg} =0$), while $\Omega=\frac{2 \mu E_T}{\hbar}$ is the Rabi 
frequency, $v_z$ is the atomic velocity along the propagation axis of the light (and 
therefore $kv_z$ is the Doppler shift). To calculate $\mathcal{L}(z)$ we have to analyse 
the atomic dynamics \cite{ducloy_jphysique_1991}. In the case of atoms moving away from 
the interface ($v_z>0$) we have $z_o=0$ because of the assumption that optical coherence 
is destroyed after a collision with the wall \cite{DUCLOY1993336}. For atoms moving 
towards the interface ($v_z<0$) $z_o \rightarrow \infty$. The contribution of arriving 
and departing atoms has been shown to be identical 
\cite{schuurmans_1976,ducloy_jphysique_1991}. Hence, we may limit our calculations to atoms 
with non-negative velocities that depart from the interface (window), simply multiplying 
the result of Eq.~(\ref{eq:sigma_eg}) by a factor of 2. 

When atom-surface interactions are present, both the atomic resonance frequency 
$\Tilde{\omega}_o(z)$ and the transition linewidth $2 \pi \Tilde{\Gamma}(z)$ are 
distance-dependent and the function $\mathcal{L}(z)$ can be written as   
\begin{equation}
\label{eq:lfunction}
    \mathcal{L}(z') - \mathcal{L}(z) = \int\limits_z^{z'} \mathrm{d}\xi\, \left[ 
    \frac{2 \pi \Tilde{\Gamma}(\xi)}{2} - \mathrm{i}\left( \omega - \Tilde{\omega_o}(\xi) 
    - kv_z\right) \right].
\end{equation}
The distance dependence of the atomic resonance frequency $\Tilde{\omega_o}(z)$ takes the 
form $\Tilde{\omega}_0(z) =\omega_0 - 2 \pi C_3/z^3$ due to non-retarded (electrostatic)
Casimir-Polder interactions. Although the effects of retardation have been computed in 
selective reflection spectra within the bounds of the infinite Doppler approximation 
\cite{carvalhoPRA2018}, for the purposes of this article we will assume non-retarded atoms 
with $C_3$ being the well-known van der Waals coefficient. The transition linewidth can 
have a similar distance dependence in the non-retarded limit, when atomic dipole 
transitions coincide with surface polariton frequencies 
\cite{hinds_atoms_1997, failacheprl1999, laliotisPRA2015}. In this case, 
$\Tilde{\Gamma}=\Gamma + \frac{\Gamma_3}{z^3}$ where $\Gamma$ includes both the natural 
linewidth as well as pressure broadening, and $\Gamma_3$ is a parameter proportional to the 
imaginary part of the image coefficient $\frac{\epsilon-1}{\epsilon+1}$ 
\cite{Acta, wylie&SipePRA1984, wylie&SipePRA1985}. This distance dependence of the 
transition linewidth, introduced for practical purposes as an 
imaginary part of the $C_3$ coefficient, has been experimentally demonstrated for flat 
surfaces \cite{failache2002}, but also for nanostructured surfaces with resonances tuned at 
optical wavelengths \cite{laliotis_scienceadv2018}. However, in most experimental 
scenarios, $\Gamma_3$ remains very small, because the dominant atomic transitions do not 
coincide with polariton resonances. Therefore, the linewidth is defined primarily by the 
vapor pressure and can be assumed to be constant. The effects of a divergent linewidth will 
be examined in the last section of this paper. Using the laser detuning, $\delta = \frac{\omega - \Tilde{\omega}_0}{2\pi}$, we can write the susceptibility as
\begin{widetext}
     \begin{equation}
     \label{eq:susceptibility}
         \Bar{\chi}(\delta) = \frac{2 N \mu^2 k}{\epsilon_o \hbar}\int_0^\infty \mathrm{d}v_z\, \frac{W(v_z)}{v_z} \int_0^\infty \mathrm{d}z \int_0^z \mathrm{d}z'\, \mathrm{e}^{2\mathrm{i}kz}\mathrm{exp}\left[ \frac{2 \pi}{v_z}\left\lbrace \left( \frac{\Gamma}{2} -\mathrm{i}\left(\delta - \frac{v_z}{\lambda}\right) \right)(z'-z) -\frac{\mathrm{i}C_3}{2}\left(\frac{1}{z'\,^2} - \frac{1}{z^2} \right) \right\rbrace \right].
     \end{equation}
\end{widetext}
 
We will denote the triple integral in Eq.~(\ref{eq:susceptibility}), onto which we will 
focus our computational efforts, as $I_{SR}(\delta)$. Finally, we will define the 
normalised selective reflection signal, $S_{SR}(\delta)$, as
\begin{equation}
\label{eq:SRsignal}
S_{SR}(\delta)= \frac{R(\delta)-R_o}{R_o} = -\frac{4 N \mu^2 k n}{\epsilon_o \hbar (n^2-1)} 
\Re [I_{SR}(\delta)]
\end{equation}
The integral $I_{SR}(\delta)$ provides the lineshape of the selective reflection signal, 
and the multiplicative factor allows us to predict its absolute amplitude. 

In most practical cases, a frequency modulation (FM) is applied to the laser frequency, 
$f_L$, whose time dependence becomes $f_L=f+Mcos(\omega_{FM} t)$, where $M$ and 
$\omega_{FM}/2\pi$ are the amplitude and frequency of the FM, respectively, and $f$ is the 
central frequency of the laser. The complete equation for the detected $S_{FMSR}$ signal is 
given in the Appendix. Recall that in the small modulation depth limit ($\Gamma \gg M$), 
$S_{FMSR}$ is proportional to the derivative of the direct signal given by
\begin{equation}
\label{eq:FMSR}
S_{FMSR}(\delta)= M \frac{d S_{SR}(\delta)}{d\delta}.
\end{equation}

In the infinite Doppler approximation, the Maxwell-Boltzmann velocity distribution is 
approximated to be constant, $W(v_z) \approx v_p \sqrt{\pi}$, where $v_p$ is the most 
probable velocity \cite{ducloy_jphysique_1991}. This only allows one to calculate the 
derivative of the selective reflection signal. Therefore the FMSR signal can only be computed in the small modulation approximation of Eq.~(\ref{eq:FMSR}), while signal distortions due to a finite FM amplitude cannot be included. In an experimental process, as presented in Refs.~\cite{chevrollier_high_1992, laliotisnatcommun2014, failacheprl1999, laliotisAPB2008, oriaepl1991, biplab_dutta_2023}, the measured selective reflection lineshapes are compared (fitted) to the results of Eq.~(\ref{eq:FMSR}), using the homogeneous linewidth $\Gamma$ and the $C_3$ coefficient as free parameters, while adjustments of a small frequency shift, offset and overall signal amplitude are also permitted. The overall quality of the fits and the statistical uncertainty of the $C_3$ measurements depends on the signal-to-noise ratio of the experiment. However, systematic errors can also critically depend on the model and the corresponding approximations used to calculate the theoretical spectra.  

The analysis that we will describe below allows us to calculate the full FM 
[Eq.~(\ref{eq:SRspectra})] or direct selective reflection signals 
[Eq.~(\ref{eq:SRsignal})], thereby eliminating a source of systematic effects on the 
experimental measurement of the $C_3$ coefficient and allowing a better comparison between theoretical and experimental spectra.

\section{Numerical calculation of selective reflection spectra}
In order to solve the triple integral numerically efficiently, we separate the 
$z$- and $z'$-dependent contributions as
\begin{equation}
\label{eq:double_integral}
    \begin{aligned}
        \int_0^\infty \mathrm{d}z \,\mathrm{e}^{2\mathrm{i}kz} \mathrm{exp}\left[ \frac{2 
        \pi}{v_z}\left\lbrace \left( \mathrm{i}\left(\delta - \frac{v_z}{\lambda}\right)- 
        \frac{\Gamma}{2}\right)z +\frac{\mathrm{i}C_3}{2z^2} \right\rbrace \right]\\
        \times \int_0^z \mathrm{d}z'\, \mathrm{exp}\left[ \frac{2 \pi}{v_z}\left\lbrace 
        \left( \frac{\Gamma}{2} - \mathrm{i}\left(\delta - \frac{v_z}{\lambda}\right)\right)z' -
        \frac{\mathrm{i}C_3}{2z'\,^2} \right\rbrace \right].
    \end{aligned}
\end{equation}
In this form, the inner integral depends on $z$ only through the upper limit of 
integration. To reduce the computational load, we define a cutoff $L$ for the $z$-integral 
and use a common step size for both the $z$- and the $z'$-integrations. This allows us to 
use a cumulative numerical integration to calculate the $z'$-integral followed by standard 
trapezoid integrations to calculate the $z$- and $v_z$-integrals. The process is then 
repeated to obtain the selective reflection spectrum for all detunings. 

The cutoff point $L$ of the integral of Eq.~\ref{eq:double_integral}) is a critical 
parameter for achieving fast numerical convergence. For this purpose, we assume a small 
vapor-induced absorption \cite{schuurmans_1976, ducloy_jphysique_1991} that exponentially 
reduces the contribution of the inner parts of the cell. We verify that the integral 
converges to a value that is not influenced by the parameters of the decay function to 
within one percent. We have tested different cutoff functions, other than a simple 
exponential, to achieve this slow fading out of the integrals. In this work we use an 
appropriately scaled and shifted logistic function of the form 
$f_\text{log}(z) = (1+e^{-k_c(z_c-z)})^{-1}$, where $z_c$ determines the point around 
which the function fades out and $k_c$ the steepness of the fading.

\begin{figure}[h]
\centering
\includegraphics[width=80mm]{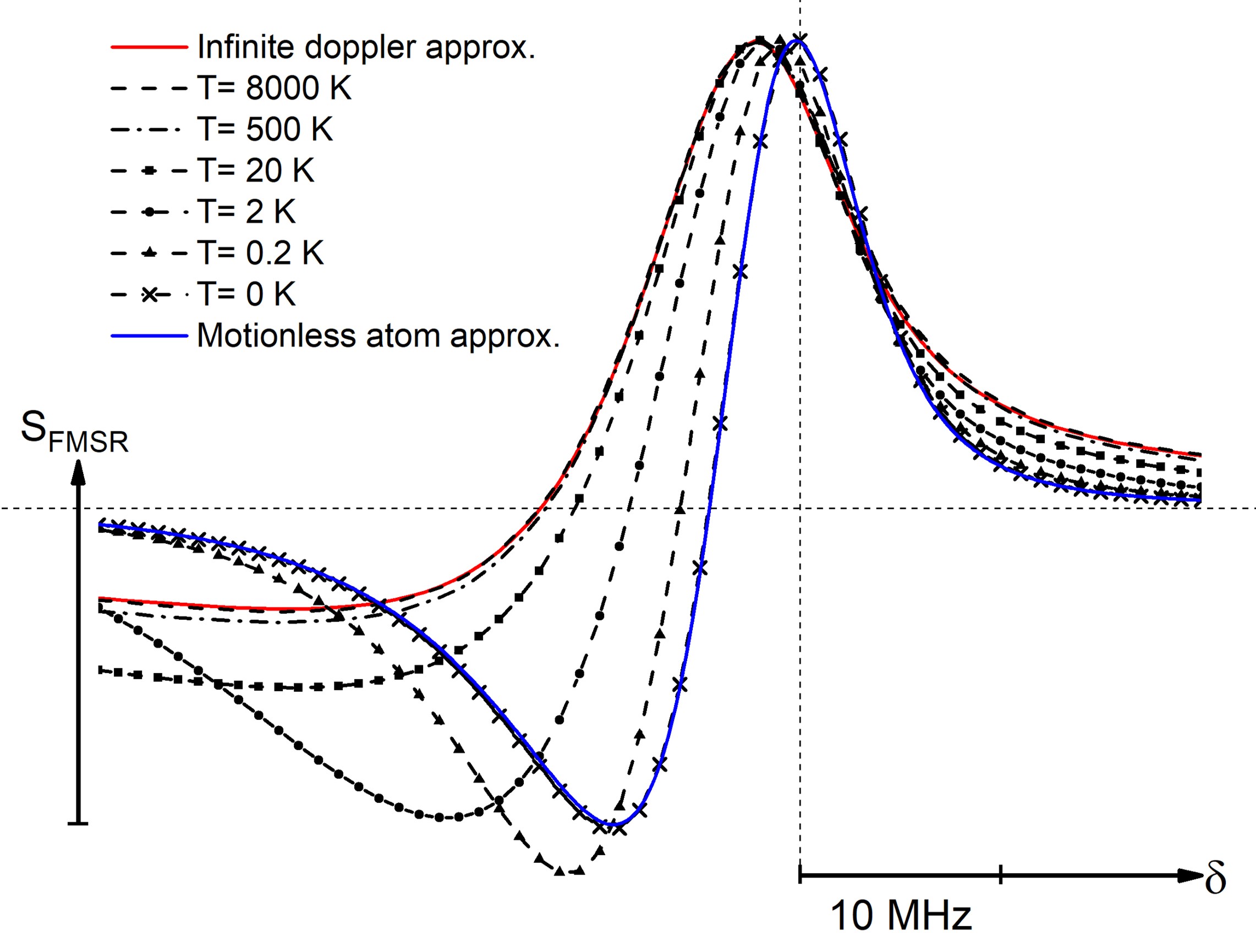}%
\caption{ $S_{FMSR}$ signal (see Eq.~(\ref{eq:FMSR})) versus detuning for different values of Maxwell-Boltzmann (MB) temperature. The amplitudes of the curves are normalised to their maximum value for a better comparison. The red and blue curves represent the infinite Doppler approximation (T$\rightarrow \infty$) and the motionaless atom approximation respectively. The spectra calculated using a Maxwell-Boltzmann velocity distribution (black lines) tend towards the blue curve for small temperatures and towards the red curve for large temperatures, demonstrating the validity of our numerical calculations. We stress that the temperature of 8000 K is chosen for illustrative purposes and does not correspond to a value that can be achieved in a vapor cell experiment. The 
van der Waals coefficient is $C_3=\rm{10~kHz~\mu m^3}$, the homogeneous width 
$\Gamma=\rm{15~MHz}$, and the wavelength is $\lambda \rm{=672~nm}$. The vertical line 
indicates the zero detuning position with respect to the resonance in the volume (away from 
the surface).
\label{Fig1}}
\end{figure}

To demonstrate the validity of our numerical results we show in Fig.~\ref{Fig1} the $S_{FMSR}$ signal, proportional to the first derivative of the selective reflection signal, for different 
translational temperatures of the vapor when the FM amplitude remains small compared to the 
homogeneous linewidth ($M \ll \Gamma$). We have chosen $C_3$=10~kHz $\rm\mu m^3$, 
$\Gamma$=15~MHz and $\lambda$=672~nm that are typical experimental values expected for 
selective reflection experiments performed on low-lying excited states of alkali atoms. 
The red curve in Fig.~\ref{Fig1} shows the derivative of the selective reflection signal 
as calculated within the infinite Doppler approximation (i.e infinite translational 
temperature) \cite{ducloy_jphysique_1991}, while the blue curve represents an analytical 
calculation performed for the case of motionless atoms. The exact numerical calculations 
tend towards the blue curve when $T\rightarrow 0$ and towards the red curve when 
$T\rightarrow \infty$. For a typical temperature in SR experiments of $T$=500~K, the 
infinite Doppler approximation matches the exact calculations rather well, thus confirming 
the analyses performed in previous experiments 
\cite{laliotisnatcommun2014, oriaepl1991, failacheprl1999}.
Figure~\ref{Fig1} validates our numerical calculations that we can now use in order to 
analyse the regime of strong atom-surface interactions (with Rydberg atoms) and quantify 
any systematic effects introduced in previous experiments.

\section{Results and Discussion}
\subsection{Rydberg-surface spectroscopy}
Having established the validity of our numerical tool, we can proceed to explore, as a 
first step, the effects of giant Casimir-Polder interactions on selective reflection 
spectra \cite{biplab_dutta_2023}. This scenario is relevant for experiments with highly 
excited Rydberg atoms that are important for interfacing atoms with photonic platforms, 
predominately for the fabrication of quantum devices and sensors, but also for fundamental 
physics applications, such as the study of multi-polar effects 
\cite{LaliotisPhysRevResearch2024}. We focus on relatively low-lying Rydberg states of 
$n^{\star}\approx$ 15 with a $C_3$ coefficient on the order of 10~MHz~$\mu m^3$, already 
performed with selective reflection and thin cell spectroscopy \cite{biplab_dutta_2023}. 

Rydberg atoms can be spectroscopically addressed via a two-step excitation, by first 
pumping to the $6P_{1/2}$ cesium level and subsequently probing of Rydberg 
transitions $6P_{1/2} \rightarrow nS, nD$, typically at wavelengths of $\approx$510~nm. 
Rydberg levels have long lifetimes when isolated. However, in a dense vapor environment 
required for selective reflection spectroscopy, the lifetimes reduce significantly due to 
collisions, leading to transition linewidths ($\Gamma$) that can be on the order of 
100~MHz.

\begin{figure}[h]
\centering
\includegraphics[width=85mm]{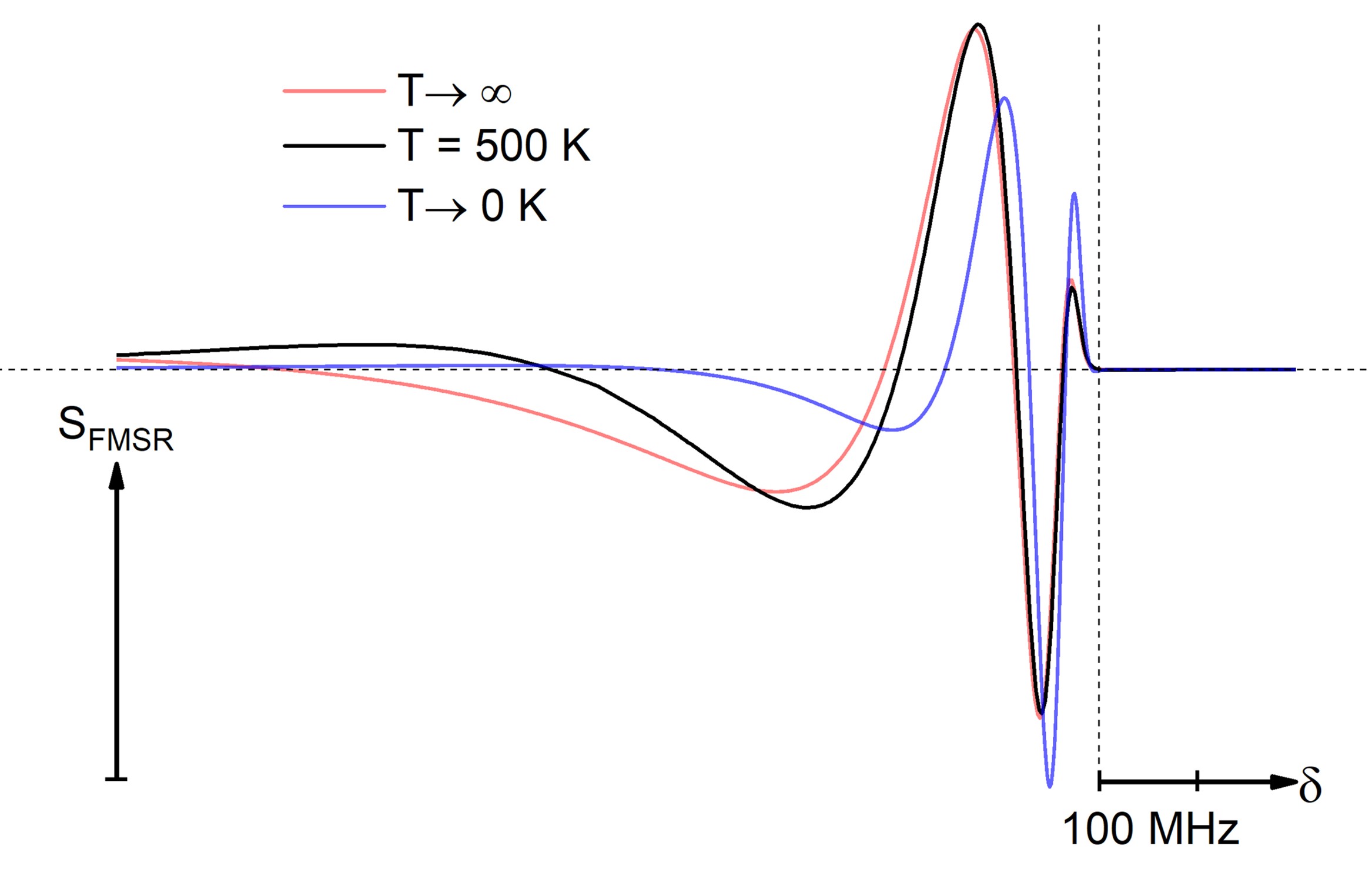}%
\caption{$S_{FMSR}$ signal in arbitrary units versus detuning on the cesium $6P_{1/2} \rightarrow 17D_{3/2}$ transition at 
$\lambda$=512~nm. The van der Waals coefficient is $C_3=\rm{8.8~MHz~\mu m^3}$ and the  
linewidth is $\Gamma=\rm{50~MHz}$. The full numerical calculation assuming a 
Maxwell-Boltzmann velocity distribution of temperature T=500 K is represented by the solid 
black line. The red and blue spectra correspond to calculations performed under the 
infinite Doppler and the motionless atom assumptions, respectively. The relative curve amplitudes are not normalized.  
\label{Fig2}}
\end{figure}

In Fig.~\ref{Fig2} we show the simulated FMSR spectra ($S_{FMSR}$) on the cesium
$6P_{1/2} \rightarrow 17D_{3/2}$ transition at 512~nm, with a $C_3$ coefficient of 
8.8~MHz~$\rm\mu m^3$. We consider here a linewidth of $\Gamma$=50~MHz corresponding to a 
cesium vapor pressure of about 15~mTorr typically used in selective reflection experiments. 
Figure~\ref{Fig2} shows the FMSR spectra calculated with our numerical model for a 
translational temperature of T=500~K (black line) that can be compared to the FMSR spectra 
calculated under the infinite Doppler approximation of Ref.~\cite{ducloy_jphysique_1991}, 
including the Doppler correction introduced in Ref.~\cite{Papag1994} (red line). 
Figure~\ref{Fig2} shows clearly that the two curves have significantly different lineshapes with different symmetries, in particular in the wings of the spectra. This clearly suggests that the infinite Doppler approximation is inadequate for the interpretation of FMSR spectra with large Casimir-Polder interactions. This is because the Casimir-Polder shifts are no longer small compared to the Doppler shift, thus strongly influencing the wings of the spectra. The above conclusions are corroborated from the analysis of experimental selective reflection spectra obtained on the $6P_{1/2} \rightarrow 16S_{1/2}$ transition reported in Ref.~\cite{biplab_dutta_2023}. These preliminary results demonstrate that the infinite Doppler approximation model cannot reproduce accurately the selective reflection spctra of Rydberg atoms.    

For comparison, we also include in Fig.~\ref{Fig2} the FMSR spectra assuming that atoms are motionless (blue line). This approximation, more appropriate for a situation in which the homogeneous linewidth is much larger than 
the Doppler width, also fails to reproduce the FMSR spectra calculated with the exact 
model (black line). It is worth mentioning that fitting the FMSR spectrum calculated with 
our exact model (black) line with the infinite Doppler approximation model, leaving $C_3$ 
and $\Gamma$ as free parameters, gives a $C_3\approx$ 5~MHz~$\mu m^3$ which is by almost 
a factor of 2 smaller than the expected value of 8.8~MHz$~\mu m^3$ that was used in our 
simulations. Additionally, the quality of the resulting fits remains limited. The above 
clearly shows that the infinite approximation model introduces important systematic errors 
and cannot be used for studying Casimir-Polder interactions with highly excited atoms. 

We also examined the possibility of using selective reflection spectroscopy for probing 
contributions of higher-order multipoles such as quadrupole-quadrupole terms to the 
Casimir-Polder interaction. It was demonstrated in 
Refs.~\cite{crossePRA2010, LaliotisPhysRevResearch2024} that, in the near field, 
quadrupole-quadrupole and dipole-octupole interactions follow a $C_5/z^5$ distance 
dependence, with detailed calculations for the $C_5$ coefficients given in 
Ref.~\cite{LaliotisPhysRevResearch2024}. Our studies show that the inclusion of the 
quadrupole coefficient $C_5$ does not substantially change the form or amplitude of the 
selective reflection spectra, even when no frequency modulation is applied. We believe 
this is because the typical probing depth of selective reflection can be pushed further 
away from the surface as the Casimir-Polder interactions increase. For example, the 
argument, that a characteristic probing depth will depend on the quantity 
$\sqrt[3]{C_3/\Gamma}$ where the Casimir-Polder shift equals the transition linewidth, 
indicates that the probing can extend several hundreds of nanometers inside the vapor, 
where the effects of quadrupole interactions will be significantly diminished. This 
clearly suggests that thin-cell spectroscopy, for which the probing depth is fixed by 
the thickness of the cell and remains independent of the Casimir-Polder strength, is 
much more favorable for probing short range higher-order effects 
\cite{LaliotisPhysRevResearch2024}. 

\subsection{Casimir-Polder spectroscopy of low-lying excited states}
It is also interesting to establish the extent and sign of systematic errors that could 
have been introduced in previous Casimir-Polder selective reflection measurements with 
low-lying excited states. We chose to analyse selective reflection spectra on the first 
cesium resonance that has been previously probed in various experiments 
\cite{laliotisAPB2008, todorovSPIE2019, oriaepl1991, laliotis_scienceadv2018}. For 
simplicity, we focus on the D1 line, whose hyperfine components are well isolated 
(approximately 1.1~GHz spacing) making the spectra easier to analyse. 

\begin{figure}[h]
\centering
\includegraphics[width=75mm]{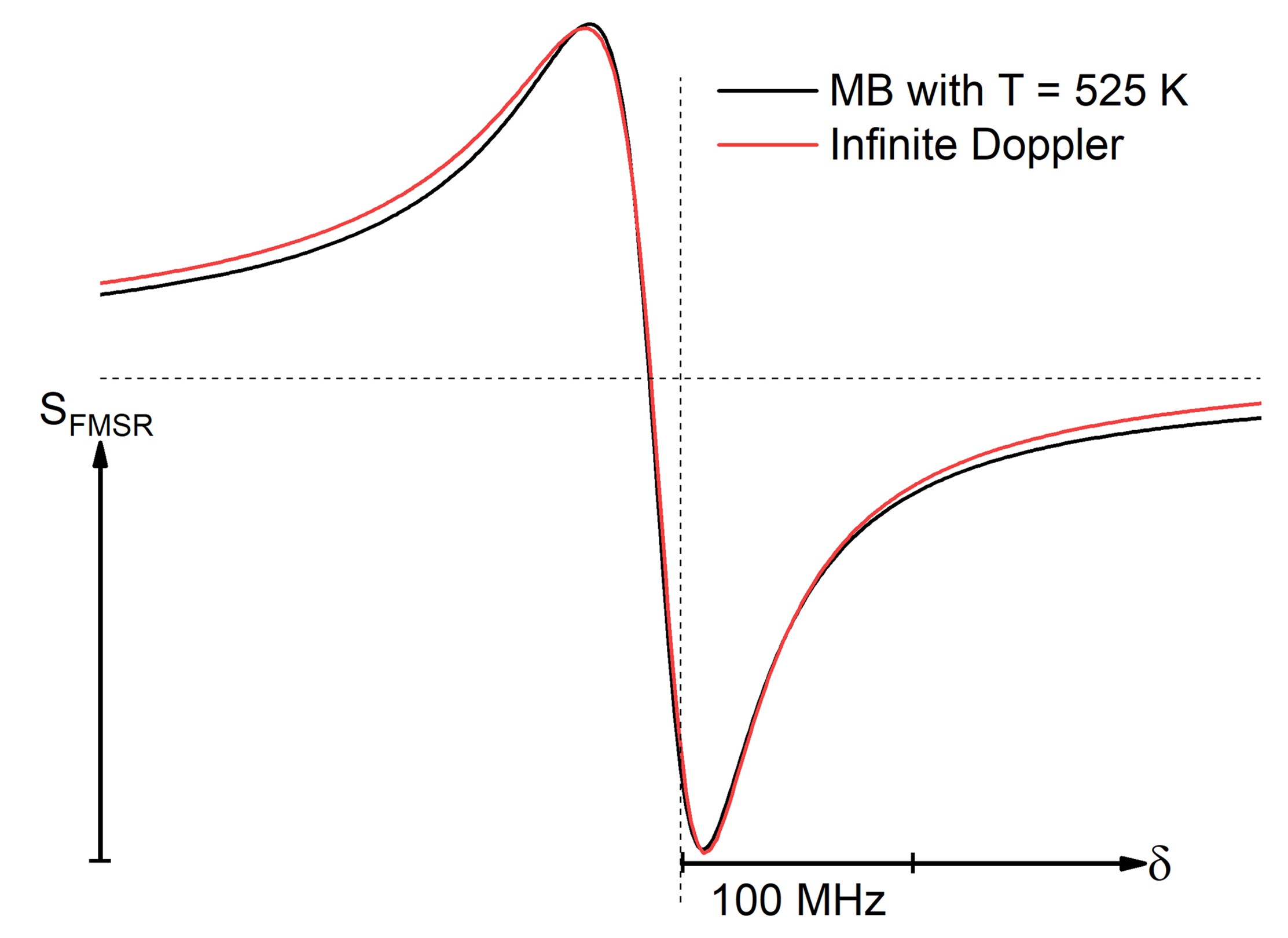}%
\caption{$S_{FMSR}$ versus detuning on the $6S_{1/2} \rightarrow 6P_{1/2}$ transition at 
$\lambda$=894 nm with a linewidth of $\Gamma$=10 MHz, corresponding to a cesium pressure 
of about 5 mTorr.  The full numerical calculation assuming a Maxwell-Boltzmann velocity 
distribution of temperature T=525 K is represented by the solid black line. The red 
spectrum corresponds to a calculation performed under the infinite Doppler approximation 
(adding the Doppler correction described in \cite{Papag1994}). The differences between the 
two curves are small but evident. This suggests that using the infinite Doppler 
approximation could lead to systematic errors in experimental spectroscopic measurements of 
the Casimir-Polder interaction.   }
\label{Fig3}
\end{figure}

In Fig.~\ref{Fig3} we show the simulated FMSR spectrum ($S_{FMSR}$) on the D1 line 
($\lambda$=894~nm) at a temperature of T=525~K, corresponding to a Doppler width of 
approximately 250~MHz. We chose a homogeneous linewidth of $\Gamma$=10~MHz, typically 
observed at low cesium vapor pressures. We note that the pressure broadening on this line 
is measured to be approximately 1~GHz/Torr \cite{laliotisAPB2008}. The spectroscopic $C_3$ 
coefficient is assumed to be 1.2~kHz~$\mu m^3$, close to its theoretically calculated value 
reported in \cite{carvalhoPRA2018, peyrotpra2019}.  Fig.~\ref{Fig3} also shows the 
selective reflection spectrum, using the same $C_3$ and $\Gamma$ parameters calculated 
within the infinite Doppler approximation. Although the two curves are similar, some 
differences are evident, suggesting that systematic errors can be introduced when analyzing 
the results of these experiments with the approximations of 
Ref.~\cite{ducloy_jphysique_1991}. We can estimate these systematics by fitting the full 
numerical spectrum with our library of curves (calculated using the infinite Doppler 
approximation), typically used in the past to analyze selective reflection experiments. Our 
analysis shows that the infinite Doppler approximation slightly underestimates the $C_3$ 
coefficient introducing errors of approximately 10\%. For the full numerical curve of 
Fig~\ref{Fig3}, for example, the $C_3$ coefficient extracted by our fits is $\approx$ 
0.95~kHz~$\mu m^3$, with a $\Gamma=$10.6~MHz. This suggests that the full numerical model 
is preferable for analyzing selective reflection experiments even when the $k u_p/\Gamma$ parameter (Doppler width divided by the natural linewidth) \cite{laliotisAPB2008} remains modest. 

\subsection{Distance-dependent lifetime}
We also use our numerical model to explore the effects of distance-dependent lifetimes on 
FMSR spectra. The modification of the atomic lifetime in the vicinity of the surface 
arises due to the existence of surface polariton modes and follows a $z^{-3}$ dependence 
in the non-retarded limit \cite{carvalho_prl2023,failache2002, ribeiro2013}. Retardation can also lead to 
distance-dependent lifetimes due to cavity QED effects \cite{SandoghdarPRA1996} and due 
to atomic emission into evanescent waves 
\cite{LukoszJOSA1977, hinds_atoms_1997, laliotisPRA2015} that can be explored by measuring 
the fluorescence at an angle larger than the angle of total reflection \cite{Burgmans1977}. 
However, such phenomena will be ignored here because they remain negligible compared to 
the strongly divergent terms introduced by the atom-polariton coupling. 

Assuming one dipole coupling, 
$\ket{e}\rightarrow \ket{a}$ (starting form the excited state), that is resonant with the 
surface polariton modes we can write the imaginary part of the $C_3$ coefficient 
$C_3^{im}$, as \cite{failache2002, carvalho_prl2023} :
\begin{equation*}
     C_3^{im} = 2 \mu_{ea} \Im\left[
     \frac{\epsilon(\omega_{ea})-1}{\epsilon(\omega_{ea})+1 }\right] 
     \frac{1}{e^{\frac{\hbar \omega_{ea}}{k_B T}}-1} .
\end{equation*}
Here we denote the dipole moment matrix element of the coupling as $\mu_{ea}$ and the 
frequency as $\omega_{ea}$, whereas $k_B$ is the Boltzmann constant and $T$ is the 
temperature of the vacuum considered to be in equilibrium with the surface. 

\begin{figure}[h]
\centering
\includegraphics[width=85mm]{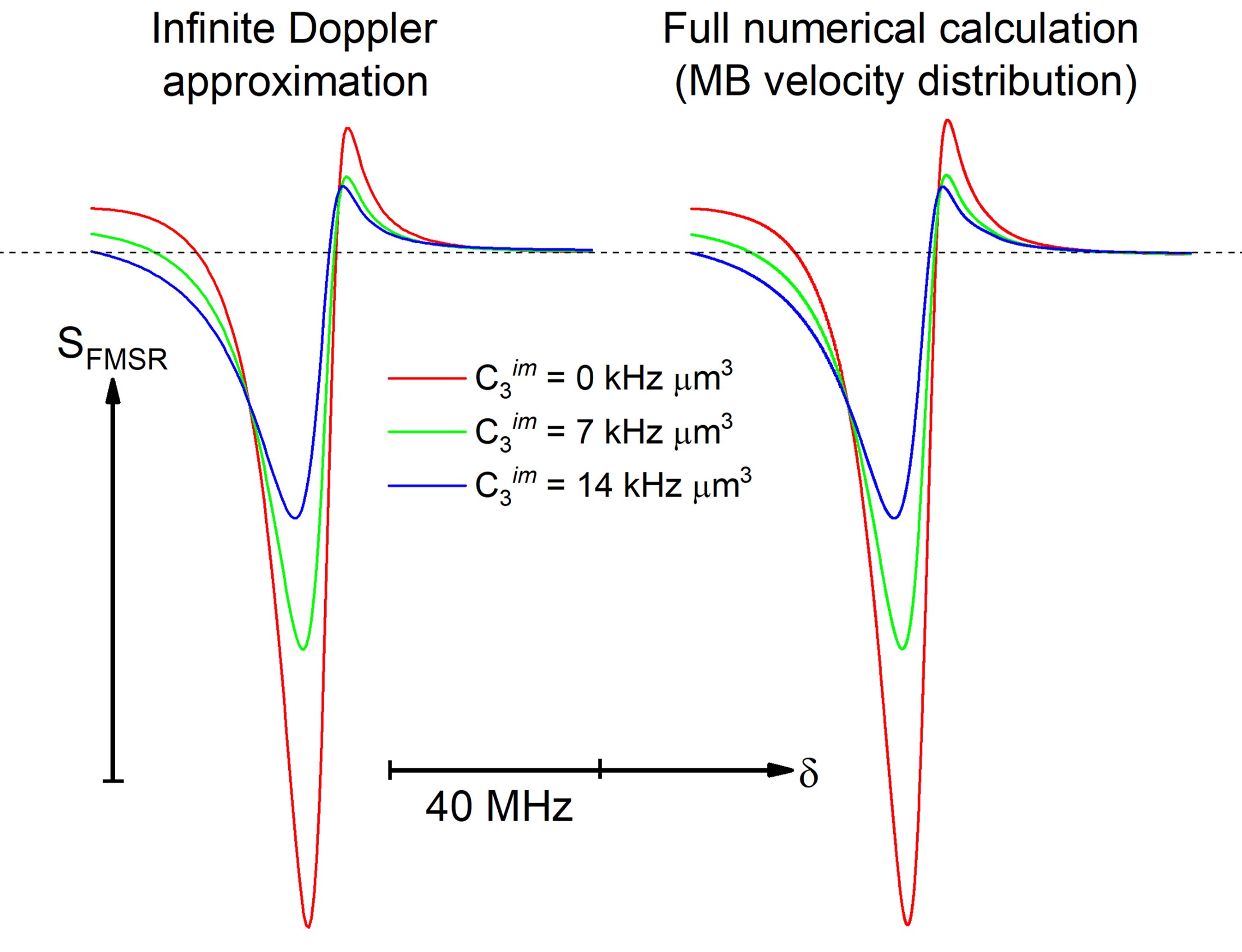}%
\caption{$S_{FMSR}$ versus detuning at cesium $\lambda$=459~nm with a linewidth of $\Gamma$=10~MHz, corresponding to the $6S_{1/2} \rightarrow 7P_{1/2}$ transition. The real part of the $C_3$ coefficient is 
14~kHz~$\mu m^3$, while the imaginary part varies between 0 (red line) to 7~kHz~$\mu m^3$ 
(green line) and 14~kHz~$\mu m^3$ (blue line). In the right panel we show the calculations 
according to our numerical model that accounts for the Maxwell-Boltzmann velocity 
distribution at a finite temperature (T=500~K), while curves calculated with the 
infinite Doppler approximation ($T\rightarrow \infty$) are shown on the left. 
}
\label{Fig4}
\end{figure}

It has been demonstrated that, using the infinite Doppler approximation, the effects of 
$C_3^{im}$ have only little influence on the shape, impacting primarily the amplitude of 
FMSR spectra \cite{carvalho_prl2023}. In Fig.~\ref{Fig4} we show the FMSR spectra with 
parameters corresponding to the $6S_{1/2} \rightarrow 7P_{1/2}$ transition of cesium at 
459~nm, experimentally explored in Ref.~\cite{carvalho_prl2023}. The $C_3$ coefficient of 
Cs($7P_{1/2}$) against a sapphire surface depends on temperature 
\cite{laliotisnatcommun2014, carvalho_prl2023} and on the exact position of the surface 
polariton frequency with respect to the strong dipole coupling 
$7P_{1/2} \rightarrow 6D_{3/2}$ at 24.7~THz. Here, in order to examine the effects of the 
distance-dependent lifetime on FMSR spectra, we will assume that the real part of the 
$C_3$ coefficient is 14~kHz~$\mu m^3$ (corresponding to its calculated value close to 
room temperature), while letting the imaginary part vary between 0, 7 and 14~kHz~$\mu m^3$. 
This is done first under the infinite Doppler approximation (on the left of 
Fig.~\ref{Fig4}) and then using our full numerical calculations (on the right of 
Fig.~\ref{Fig4}). It is clear that both models give very similar results and the infinite Doppler approximation accurately predicts 
the evolution of the signal amplitude with increasing $C_3^{im}$. This suggests that the 
disagreements between theory and experiment in \cite{carvalho_prl2023}, concerning the 
predictions of the relative amplitude between the $6S_{1/2} \rightarrow 7P_{1/2}$ and the 
$6S_{1/2} \rightarrow 7P_{3/2}$ spectra, are not amenable to systematics introduced by the 
infinite Doppler approximation that was used to analyse the spectra in Ref.~\cite{carvalho_prl2023}. 

\section{Conclusions}

We presented full numerical solutions of selective reflection integrals allowing one
to account for the Maxwell-Boltzmann velocity distribution of atoms without having to 
assume that the Doppler width is infinitely larger than the homogeneous linewidth. We have
shown that the full numerical model is necessary for interpreting atom-surface interaction 
measurements with highly excited Rydberg atoms where the Casimir-Polder interactions 
become sufficiently large to influence the wings of the selective reflection 
spectra \cite{biplab_dutta_2023}. Furthermore, we demonstrated that the model allows eliminating systematic errors on the order of 10 \% in atom-surface interaction measurements for low-lying excited atoms. 

The new numerical model could 
therefore find use in analyzing a new generation of selective reflection measurements 
including electric-dipole forbidden transitions \cite{Chan2025}, interactions 
of atoms with meta-materials \cite{laliotis_scienceadv2018} and interactions of Rydberg 
atoms with dielectric surfaces. Our model could also lead to better agreement between 
experiment and theory for selective reflection experiments with magnetic fields 
\cite{papageorgiou_APB_1994} or selective reflection at high densities to demonstrate 
cooperative phenomena such as the Lorentz-Lorenz shift 
\cite{wang_pra_1997, guo_pra_1996, VULETIC1993} and the cooperative Lamb shift 
\cite{Friedberg1973,Friedberg2022}, that has been experimentally studied with thin cell 
spectroscopy \cite{KeaveneyPRL2012, PeyrotPRL2018,Dobbertin2020}. Finally, the analysis presented here provides insights into the interpretation of the spectroscopic response of atoms close to surfaces for one-photon or two-photon thin-cell experiments \cite{fichet_epl_2007, biplab_dutta_2023, kublernatphot2010}, as well as Electromagnetically Induced Transparency (EIT) experiments on Rydberg atoms close to surfaces \cite{sedlacek_microwave_2012, schlossberger_rydberg_2024} that can be used for measuring electric fields at microwave or even THz frequencies.

\acknowledgments
This work was financially supported by the ANR-DFG 
grant SQUAT (Grant Nos. ANR-20-CE92-0006-01 and DFG SCHE 612/12-1), the DAAD and Campus 
France (via the PHC-PROCOPE programme, grant Nos. 57513024 and 57733761), and the French 
Embassy in Germany (via the Campus-France PHC-Procope project 44711VG and via PROCOPE 
Mobilité, project DEU-22-0004 LG1).

\appendix*
\section{Frequency modulation}
When a frequency modulation (FM) is applied, the time dependence of the laser frequency 
becomes $f_L=f+M\cos(\omega_{FM} t)$, where $M$ and $\omega_{FM}/2\pi$ are the amplitude 
and frequency of the FM, respectively, and $f$ is the central frequency 
of the laser. In this case, the in-phase component of the normalised selective reflection 
signal oscillating at frequency $f_{FM}$, denoted as $S_{SRFM}$, is given by
\begin{widetext}
\begin{equation}
\label{eq:SRspectra}
S_{FMSR}(\delta)= -\frac{4 N \mu^2 k n}{\epsilon_o \hbar (n^2-1)} \Re \left[ \sum_N \left[ 
I_{SR}(\delta+N \omega_{FM})+ I^{\ast}_{SR}(\delta+N \omega_{FM}-\omega_{FM}) \right] 
J_N\left(\frac{M}{\omega_{FM}}\right) J_{N-1}\left(\frac{M}{\omega_{FM}}\right) \right],
\end{equation}
\end{widetext}
where $J_N(x)$ is the $N$th order Bessel function of the first kind.

\bibliographystyle{apsrev4-1}
\bibliography{biblioMoleculesvBD}

\end{document}